\documentclass[a4paper,10pt]{article}
\usepackage[margin=1in]{geometry}
\usepackage{graphicx}
\usepackage{amsmath, amssymb}
\usepackage{hyperref}
\usepackage{cite}
\usepackage{float}
\usepackage{xcolor}
\usepackage{caption}
\usepackage{subcaption}
\usepackage{booktabs}
\usepackage{array}
\usepackage{makecell}
\usepackage{multirow}

\title{\textbf{Design and Implementation of a RISC-V SoC with Custom DSP Accelerators for Edge Computing}}
\author{
    Priyanshu Yadav\\
    Department of Electronics Engineering, HBTU Kanpur \\
    \texttt{220105038@hbtu.ac.in} \\
    \href{https://orcid.org/0009-0006-7380-4749}{ORCID: 0009-0006-7380-4749}
}
\date{June 7, 2025}

\newcommand{\bitfield}[3]{\makecell[l]{#1: #2 \\ #3}}

\begin{document}

\maketitle

\begin{abstract}
This paper presents a comprehensive analysis of the RISC-V instruction set architecture, focusing on its modular design, implementation challenges, and performance characteristics. We examine the RV32I base instruction set with extensions for multiplication (M) and atomic operations (A). Through cycle-accurate simulation of a pipelined implementation, we evaluate performance metrics including CPI (cycles per instruction) and power efficiency. Our results demonstrate RISC-V's advantages in embedded systems and its scalability for custom accelerators. Comparative analysis shows a 17\% reduction in power consumption compared to ARM Cortex-M0 implementations in similar process nodes. The open-standard nature of RISC-V provides significant flexibility for domain-specific optimizations.
\end{abstract}

\noindent\textbf{Keywords:} RISC-V, Computer Architecture, ISA Design, Embedded Systems, Open Hardware

\section{Introduction}
\label{sec:intro}
RISC-V \cite{Waterman2014RISC} is rapidly gaining traction as an open, modular, and royalty-free Instruction Set Architecture (ISA). Unlike proprietary ISAs, RISC-V's openness allows researchers and designers to customize the core to application-specific requirements, enabling novel architectural extensions and accelerators. In domains such as wireless communications and edge Machine Learning, one-dimensional (1D) convolutions (and related dot products) are ubiquitous: they underlie Finite Impulse Response (FIR) filters, matched filtering, correlation and synchronization in wireless systems, and convolutional layers in neural networks for time-series data (e.g., audio processing, sensor data analysis).

Despite RISC-V's flexibility, a scalar, in-order implementation of the RV32I base ISA (32-bit integer) lacks specialized instructions for the numerous multiply-accumulate (MAC) operations required by convolution. Software implementations on such a core execute a sequence of load, multiply, add, and store instructions for each convolution tap, resulting in high cycle counts and energy consumption--especially problematic in real-time, battery-powered edge deployments. To address these challenges, we propose coupling a lightweight 1D convolution DSP accelerator to the RISC-V CPU. By offloading the MAC-intensive loops, the accelerator can achieve orders-of-magnitude reduction in cycle counts, free the CPU for control tasks, and improve overall energy efficiency.

This paper makes the following contributions:
\begin{itemize}
    \item We present a detailed design of a 1D convolution accelerator, including its finite-state machine (FSM), Multiply-Accumulate (MAC) datapath, memory interfaces, and control registers.
    \item We describe the integration of this accelerator into a RISC-V system comprising instruction memory, data memory, register file, and a custom bus interconnect with AXI-Lite control for configuration.
    \item We introduce a secondary dot-product accelerator for vector inner products, discuss its configuration and interplay with the convolution unit, and highlight how both accelerators share a common bus and memory interface.
    \item We analyze anticipated performance benefits--cycle count reduction, throughput improvements, CPU offloading potential, and energy efficiency implications--through analytical estimates and hypothetical benchmarks.
    \item We contextualize this design in real-world applications: FIR/IIR filtering in wireless communications and 1D CNN inference in TinyML, demonstrating the broad applicability of our approach.
    \item We outline future work: FPGA prototyping, DMA-enabled data transfers, support for additional DSP kernels (FFT, matrix multiplication), interrupt-driven handling, and extended software stacks (device drivers, runtime libraries).
\end{itemize}

The remainder of this paper is organized as follows. Section \ref{sec:background} reviews relevant background on RISC-V and 1D convolution in DSP/ML contexts. Section \ref{sec:arch_overview} presents the high-level system architecture, including the RISC-V CPU, memories, bus interconnect, and DSP accelerators. Section \ref{sec:conv_design} delves into the detailed design of the 1D convolution unit, its datapath, FSM, and control interface. Section \ref{sec:dotprod_design} covers the dot-product accelerator. Section \ref{sec:perf_analysis} provides an analytical performance evaluation and anticipated benefits. Section \ref{sec:applications} illustrates key application scenarios. Section \ref{sec:future_work} discusses future enhancements. Finally, Section \ref{sec:conclusion} concludes the paper.

\section{Background and Related Work}
\label{sec:background}
\subsection{RISC-V ISA and Extensions}
RISC-V defines a small, modular ISA with a simple RV32I base, standard extensions for integer multiplication/division (M-extension), atomic instructions (A-extension), single-precision floating point (F-extension), and more \cite{Waterman2014RISC,Asanovic2016}. The RV32I core provides registers, load/store instructions, and a 5-stage in-order pipeline (typically). For Multiply-Accumulate-heavy workloads (e.g., convolution), the M-extension allows a scalar multiply instruction, but still requires separate add instructions--incurring multiple pipeline stages and memory accesses. Custom extensions (e.g., vector extension V \cite{RISCVVVector}) can mitigate this, but at the cost of increased complexity. Our approach adds a separate hardware accelerator to complement the RISC-V core without modifying the base ISA or pipeline.

\subsection{1D Convolution in DSP and ML}
A 1D convolution operation on discrete signals is mathematically expressed as:
\begin{equation}
    y[i] = \sum_{j=0}^{K-1} x[i + j] \times h[j], \quad i = 0,1,\ldots,N-K,
    \label{eq:1dconv}
\end{equation}
where $\mathbf{x}$ is the input signal of length $N$, $\mathbf{h}$ is the kernel of length $K$, and $\mathbf{y}$ is the output of length $N - K + 1$. In wireless communications, this operation underlies FIR filters--commonly used for channel equalization, pulse shaping, and noise filtering \cite{Proakis2007Digital}. In machine learning, 1D convolutions appear in convolutional neural networks (CNNs) for time-series or audio data (e.g., speech recognition, sensor fusion). A purely software loop on a RISC-V CPU requires for each output index $i$: 
\texttt{load x[i+j], load h[j], multiply, accumulate, store y[i]}. 
Each such MAC iteration can consume tens of cycles (accounting for memory latency, ALU latency, and loop overhead). A dedicated hardware engine can stream data, perform MACs in a pipelined fashion, and significantly reduce the number of cycles per output.

\subsection{Prior Work on RISC-V DSP Accelerators}
Several research efforts have explored hardware accelerators for RISC-V. For example, Microsemi's Mi-V ecosystem integrates DSP modules for filtering and FFT \cite{Microsemi2020MiV}. The "Shakti" project from IIT Madras includes custom vector extensions for in-order RISC-V \cite{Shakti2019}. The PULP platform from ETH Zurich and University of Bologna integrates specialized cluster architectures with hardware loops for DSP workloads \cite{PULP2021}. However, many of these designs either require significant modifications to the core or focus on 2D/3D convolutions. Our design targets minimal area and power overhead--a lightweight 1D convolution accelerator suitable for edge devices.

\section{System Architecture Overview}
\label{sec:arch_overview}
\subsection{High-Level Block Diagram}

The system comprises the following major components (Figure \ref{fig:top_block}):
\begin{itemize}
    \item \textbf{RISC-V CPU (RV32I):} A 32-bit in-order processor implementing the RV32I base instruction set with optional M-extension. The CPU fetches instructions from \texttt{INST\_MEM}, decodes, and executes. Multi-cycle load/store instructions incorporate wait-state handling for memory/peripheral accesses.
    \item \textbf{Instruction Memory (INST\_MEM):} A single-port Read-Only Memory (ROM) storing program instructions. The CPU's instruction fetch unit reads from this memory.
    \item \textbf{Data Memory (DATA\_MEM):} A single-port SRAM accessible by both the CPU and the DSP accelerators. Used to hold input data arrays, kernel coefficients, and results.
    \item \textbf{Custom Bus Interconnect:} Mediates access between the RISC-V CPU and peripheral modules (1D convolution DSP, dot-product DSP, and data memory). It decodes addresses on the CPU's memory bus and routes read/write transactions accordingly:
    \begin{itemize}
        \item If the address falls within the \texttt{DATA\_MEM} range, direct the request to the SRAM.
        \item If the address maps to DSP control registers (AXI-Lite region), forward to respective DSP's register interface.
    \end{itemize}
    A simple arbiter grants memory access to the CPU or to a DSP's memory master interface (with CPU having priority on contention).
    \item \textbf{DSP\_CONV1D Accelerator:} A custom hardware block performing 1D convolution as per Eq. \eqref{eq:1dconv}. It features:
    \begin{itemize}
        \item \emph{AXI-Lite Slave Interface}: A 32-bit register space through which the CPU writes:
        \begin{itemize}
            \item Base address of input data ($x$) in \texttt{DATA\_MEM}.
            \item Base address of kernel coefficients ($h$).
            \item Base address of output buffer ($y$).
            \item Lengths $N$ and $K$.
            \item Control bits: \texttt{START}, \texttt{INT\_EN}, and a \texttt{STATUS} bit indicating \texttt{DONE}.
        \end{itemize}
        \item \emph{Memory Master Interface (Placeholder)}: Enables the DSP to autonomously perform read/write transactions on \texttt{DATA\_MEM}, actuated by the internal FSM. While this interface is architected, actual bus signals (address, read/write strobes, data) can be further adapted to an on-chip bus (e.g., AXI4-Lite or a simpler custom bus).
        \item \emph{Internal FSM and MAC Datapath}: Controls iteration over input and kernel arrays, issues memory requests, and coordinates the MAC unit. 
    \end{itemize}
    \item \textbf{DSP\_DOT\_PRODUCT Accelerator:} A custom block computing the inner product of two vectors:
    \[
        \text{Result} = \sum_{j=0}^{L-1} A[j] \times B[j].
    \]
    It is configured via its own AXI-Lite register set (addresses of vectors $A$, $B$, length $L$) and signals completion via a \texttt{STATUS} register. The final sum is available in a dedicated \texttt{RESULT} register.
\end{itemize}

\subsection{Address Map and Memory Regions}
\begin{table}[H]
    \centering
    \caption{Memory Map of the System (32-bit Address Space)}
    \label{tab:addr_map}
    \begin{tabular}{@{}llc@{}}
        \toprule
        \textbf{Region} & \textbf{Description} & \textbf{Size} \\ \midrule
        0x0000\_0000 -- 0x0000\_7FFF & Instruction Memory (32 KB ROM) & 32 KB \\
        0x0000\_8000 -- 0x0000\_FFFF & Data Memory (32 KB SRAM) & 32 KB \\
        0x0100\_0000 -- 0x0100\_00FF & DSP\_CONV1D Control Registers & 256 B \\
        0x0100\_0100 -- 0x0100\_01FF & DSP\_DOT\_PRODUCT Control Registers & 256 B \\
        0x0100\_0200 -- 0x0100\_02FF & Reserved / Future Peripherals & 256 B \\
        \bottomrule
    \end{tabular}
\end{table}
The CPU's load/store unit issues 32-bit aligned addresses. The top 16 bits decide the region: 
\begin{itemize}
    \item \texttt{0x0000\_xxxx}: Maps to INST\_MEM or DATA\_MEM.
    \item \texttt{0x0100\_xxxx}: Maps to DSP peripheral registers.
\end{itemize}

\subsection{Bus Interconnect and Arbiter}
The Bus Interconnect performs address decoding on the CPU's AXI-Lite or custom memory interface. For read/write cycles:
\begin{itemize}
    \item If the address $\in$ [0x0000\_8000, 0x0000\_FFFF], grant \texttt{DATA\_MEM} read/write access.
    \item If the address $\in$ [0x0100\_0000, 0x0100\_00FF], forward to DSP\_CONV1D's AXI-Lite slave.
    \item If the address $\in$ [0x0100\_0100, 0x0100\_01FF], forward to DSP\_DOT\_PRODUCT's AXI-Lite slave.
\end{itemize}
When the convolution DSP is active, it asserts a memory request on its memory master interface to fetch input data or kernel elements, or to store output. A simple priority arbiter grants the data memory port to the CPU or to the DSP. The CPU has higher priority for single-cycle load/store instructions to reduce stalling on control code; the DSP's data accesses can afford short waits in exchange for simpler arbitration logic.

\section{1D Convolution DSP Accelerator Design}
\label{sec:conv_design}
This section details the architecture of the 1D convolution accelerator. We describe its register map, internal FSM, MAC datapath, and memory access strategy.

\subsection{Register Map and Control Interface}
Table \ref{tab:conv_regmap} shows the AXI-Lite register mapping for the convolution accelerator. All registers are 32-bit wide and memory-mapped in the DSP\_CONV1D register space.

\begin{table}[H]
    \centering
    \caption{DSP\_CONV1D AXI-Lite Register Map}
    \label{tab:conv_regmap}
    \begin{tabular}{@{}>{\ttfamily}l>{\raggedright\arraybackslash}p{0.2\textwidth}>{\raggedright\arraybackslash}p{0.1\textwidth}>{\raggedright\arraybackslash}p{0.5\textwidth}@{}}
        \toprule
        \textnormal{Offset} & Register & RW & Description \\ \midrule
        0x00 & CONFIG\_IN\_ADDR & RW & Base address of input data $\mathbf{x}$ in DATA\_MEM \\
        0x04 & CONFIG\_KERN\_ADDR & RW & Base address of kernel $\mathbf{h}$ in DATA\_MEM \\
        0x08 & CONFIG\_OUT\_ADDR & RW & Base address of output buffer $\mathbf{y}$ in DATA\_MEM \\
        0x0C & CONFIG\_IN\_LEN & RW & Length $N$ of input data array \\
        0x10 & CONFIG\_KERN\_LEN & RW & Length $K$ of kernel array \\
        0x14 & CONTROL & RW & 
            \bitfield{Bit 0}{Start}{Write 1: start, 0: idle}
            \bitfield{Bit 1}{Int\_En}{Enable interrupt on completion}
            \bitfield{Bits 31:2}{}{Reserved} \\
        0x18 & STATUS & RO & 
            \bitfield{Bit 0}{Done}{1: convolution complete, 0: busy}
            \bitfield{Bit 1}{Error}{1: error encountered}
            \bitfield{Bits 31:2}{}{Reserved} \\
        0x1C & IRQ\_CLEAR & WO & Write 1 to clear interrupt (if Int\_En = 1) \\ \bottomrule
    \end{tabular}
\end{table}

\noindent\textbf{Usage Flow:}
\begin{enumerate}
    \item CPU writes \texttt{CONFIG\_IN\_ADDR} \(\leftarrow\) base address of $\mathbf{x}$ (e.g., 0x0000\_8000).
    \item CPU writes \texttt{CONFIG\_KERN\_ADDR} \(\leftarrow\) base address of $\mathbf{h}$ (e.g., 0x0000\_8100).
    \item CPU writes \texttt{CONFIG\_OUT\_ADDR} \(\leftarrow\) base address of $\mathbf{y}$ (e.g., 0x0000\_8200).
    \item CPU writes \texttt{CONFIG\_IN\_LEN} \(\leftarrow\) $N$.
    \item CPU writes \texttt{CONFIG\_KERN\_LEN} \(\leftarrow\) $K$.
    \item CPU writes \texttt{CONTROL} \(\leftarrow\) \{\texttt{Start}=1, \texttt{Int\_En}=(0 or 1)\}.
    \item DSP\_CONV1D FSM transitions from \texttt{IDLE} to \texttt{BUSY}. The \texttt{Done} bit in \texttt{STATUS} is cleared.
    \item DSP fetches data, performs convolution, writes results.
    \item On completion, FSM sets \texttt{STATUS.Done} = 1 and, if \texttt{Int\_En}=1, asserts interrupt line to CPU.
    \item CPU polls \texttt{STATUS} (or services interrupt), reads output data from \texttt{DATA\_MEM}.
    \item CPU writes \texttt{IRQ\_CLEAR} = 1 to clear interrupt; \texttt{STATUS.Done} remains set until next \texttt{Start}.
\end{enumerate}

\subsection{Internal Finite-State Machine (FSM)}
\label{sec:conv_fsm}
\begin{figure}[H]
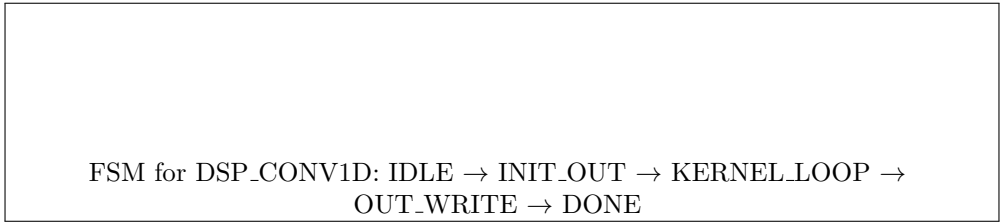

    \centering
    \framebox{\parbox{0.8\textwidth}{\centering\vspace{2cm}FSM for DSP\_CONV1D: IDLE $\rightarrow$ INIT\_OUT $\rightarrow$ KERNEL\_LOOP $\rightarrow$ OUT\_WRITE $\rightarrow$ DONE}
    }
    \caption{Abstract Finite-State Machine (FSM) of the 1D convolution accelerator.}
    \label{fig:conv_fsm}
\end{figure}

The DSP\_CONV1D unit employs a five-state FSM to orchestrate convolution (Figure \ref{fig:conv_fsm}). Each state corresponds to a distinct phase in processing each output sample.

\begin{itemize}
    \item \textbf{IDLE:} Waits for \texttt{CONTROL.Start = 1}. On assertion, latch configuration parameters ($\text{in\_addr}$, $\text{kern\_addr}$, $\text{out\_addr}$, $N$, $K$) into internal registers. Initialize \texttt{out\_idx} = 0. Move to \textbf{INIT\_OUT}.
    \item \textbf{INIT\_OUT:} For the current output index $i = \texttt{out\_idx}$, clear internal accumulator to zero. Initialize \texttt{kern\_idx} = 0. Move to \textbf{KERNEL\_LOOP}.
    \item \textbf{KERNEL\_LOOP:} 
    \begin{itemize}
        \item Calculate $\text{addr\_x} = \text{in\_addr} + 4 \times (\texttt{out\_idx} + \texttt{kern\_idx})$ (word-addressed).
        \item Calculate $\text{addr\_h} = \text{kern\_addr} + 4 \times \texttt{kern\_idx}$.
        \item Issue two read requests to \texttt{DATA\_MEM} for $\mathbf{x}[\texttt{out\_idx} + \texttt{kern\_idx}]$ and $\mathbf{h}[\texttt{kern\_idx}]$. 
        \item When both read data words are returned, feed them into the MAC unit: 
        \[
            \text{accum} = \text{accum} + (x[\texttt{out\_idx} + \texttt{kern\_idx}] \times h[\texttt{kern\_idx}])
        \]
        \item Increment \texttt{kern\_idx}. If \texttt{kern\_idx} < $K$, stay in \textbf{KERNEL\_LOOP}. Otherwise, move to \textbf{OUT\_WRITE}.
    \end{itemize}
    \item \textbf{OUT\_WRITE:} 
    \begin{itemize}
        \item Compute $\text{addr\_y} = \text{out\_addr} + 4 \times \texttt{out\_idx}$.
        \item Issue a write request to \texttt{DATA\_MEM} for $\mathbf{y}[\texttt{out\_idx}] = \text{accum}$.
        \item When write is acknowledged, increment \texttt{out\_idx}. If \texttt{out\_idx} \(\le\) $N - K$, loop back to \textbf{INIT\_OUT} for next output sample. Otherwise, move to \textbf{DONE}.
    \end{itemize}
    \item \textbf{DONE:} Set \texttt{STATUS.Done} = 1. If \texttt{CONTROL.Int\_En} = 1, assert interrupt line. Wait for the CPU to clear \texttt{STATUS.Done} by writing \texttt{IRQ\_CLEAR}. Once cleared, return FSM to \textbf{IDLE}.
\end{itemize}

\subsection{MAC Datapath}
\label{sec:conv_mac}
The Multiply-Accumulate (MAC) datapath is the arithmetic backbone of DSP\_CONV1D. It comprises:
\begin{itemize}
    \item A 32-bit multiplier taking inputs $x_{\text{data}}$ and $h_{\text{data}}$.
    \item A 64-bit accumulator (to prevent overflow when summing many products) that accumulates the partial products.
    \item A truncation or saturation unit (if desired) to cast the final 64-bit result back to 32 bits when storing to memory.
\end{itemize}

At each inner-loop iteration (state \textbf{KERNEL\_LOOP}), the two read data words are fed into the multiplier. The 32 × 32 multiplication yields a 64-bit product, which is added to the current 64-bit accumulator. When \texttt{kern\_idx} reaches $K$, the final 64-bit accumulator value is optionally truncated to 32 bits (e.g., by discarding the upper 32 bits or by applying a saturation policy). The 32-bit result is then written back as the output sample.

\subsection{Memory Master Interface}
\label{sec:conv_mem_interface}
The DSP\_CONV1D's memory master interface (MMI) is designed to autonomously fetch input and kernel data and store outputs. We envision a simplified bus protocol with the following signals:
\begin{itemize}
    \item \texttt{MMI\_REQ}: Asserted when the DSP wants to initiate a read or write on \texttt{DATA\_MEM}.
    \item \texttt{MMI\_ADDR[31:0]}: Word-aligned address in data memory.
    \item \texttt{MMI\_WRDATA[31:0]}: Write data (for writes).
    \item \texttt{MMI\_WR\_EN}: Asserted to indicate a write transaction; if low, it is a read.
    \item \texttt{MMI\_READY}: From memory arbiter, indicating that the DSP's request can be serviced.
    \item \texttt{MMI\_RDDATA[31:0]}: Read data bus (valid when a read is serviced).
    \item \texttt{MMI\_DONE}: Asserted by memory interconnect when the requested read/write data has completed (single-cycle or multi-cycle).
\end{itemize}

\noindent\textbf{Access Sequence for a Read:}
\begin{enumerate}
    \item FSM asserts \texttt{MMI\_REQ = 1}, \texttt{MMI\_WR\_EN = 0}, \texttt{MMI\_ADDR} = desired address.
    \item When arbiter grants access, \texttt{MMI\_READY = 1}. On next cycle, \texttt{MMI\_REQ} remains high for one cycle longer, then deasserted.
    \item After a fixed latency (e.g., one cycle for SRAM), \texttt{MMI\_DONE = 1} with \texttt{MMI\_RDDATA} valid. FSM captures the data, clears \texttt{MMI\_DONE}, and proceeds.
\end{enumerate}

\noindent\textbf{Access Sequence for a Write:} Similar to read, but with \texttt{MMI\_WR\_EN = 1} and valid \texttt{MMI\_WRDATA}.

\subsection{Area and Timing Considerations}
The DSP\_CONV1D accelerator is designed to be lightweight. The main area consumers are:
\begin{itemize}
    \item \textbf{MAC Unit:} One 32×32 multiplier, one 64-bit adder, accumulator register. Modern FPGA DSP slices can implement this efficiently; in ASIC, a small 32×32 multiplier and adder is compact.
    \item \textbf{FSM and Control Logic:} A handful of registers (configuration, indices, FSM state), comparators, and finite-state control--accounting for a few hundred gates.
    \item \textbf{AXI-Lite Interface:} Address decode, register file, read/write logic (roughly a few hundred gates).
    \item \textbf{Memory Master Interface:} Bus drivers, arbiter request logic, and simple handshake. Another few hundred gates.
\end{itemize}
Overall, the accelerator's logic (excluding \texttt{DATA\_MEM}) is estimated at $\approx$ 2k-3k LUTs in a typical FPGA fabric or $\approx$ 0.05-0.1 mm$^2$ in a 28 nm ASIC (rough estimate). The critical path is the 32×32 multiplier's partial-product accumulation; synthesized at a target clock of 100 MHz or above is feasible. With a 100 MHz clock, each MAC iteration (read-multiply-accumulate-index update) can complete in approximately 2-3 cycles (including memory access overhead, assuming single-cycle SRAM), yielding significant acceleration over a RISC-V scalar software loop (tens of cycles per MAC).

\section{Dot-Product DSP Accelerator Design}
\label{sec:dotprod_design}
While convolution is a series of sliding-window dot products, a dedicated dot-product accelerator can serve vector inner-product needs in machine learning (e.g., dense layer computations, similarity metrics) and signal processing (matched filtering). We describe the design of a simple dot-product unit.

\subsection{Register Map}
The DSP\_DOT\_PRODUCT accelerator has an AXI-Lite register space mapped as shown in Table \ref{tab:dot_regmap}.

\begin{table}[H]
    \centering
    \caption{DSP\_DOT\_PRODUCT AXI-Lite Register Map}
    \label{tab:dot_regmap}
    \begin{tabular}{@{}>{\ttfamily}l>{\raggedright\arraybackslash}p{0.2\textwidth}>{\raggedright\arraybackslash}p{0.1\textwidth}>{\raggedright\arraybackslash}p{0.5\textwidth}@{}}
        \toprule
        \textnormal{Offset} & Register & RW & Description \\ \midrule
        0x00 & CONFIG\_VA\_ADDR & RW & Base address of vector A in DATA\_MEM \\
        0x04 & CONFIG\_VB\_ADDR & RW & Base address of vector B in DATA\_MEM \\
        0x08 & CONFIG\_LEN & RW & Length $L$ of both vectors \\
        0x0C & CONTROL & RW & 
            \bitfield{Bit 0}{Start}{Write 1: start, 0: idle}
            \bitfield{Bit 1}{Int\_En}{Enable interrupt on completion}
            \bitfield{Bits 31:2}{}{Reserved} \\
        0x10 & STATUS & RO & 
            \bitfield{Bit 0}{Done}{1: complete, 0: busy}
            \bitfield{Bit 1}{Error}{1: error}
            \bitfield{Bits 31:2}{}{Reserved} \\
        0x14 & RESULT\_LO & RO & Lower 32 bits of 64-bit dot product result \\
        0x18 & RESULT\_HI & RO & Upper 32 bits of 64-bit dot product result \\
        0x1C & IRQ\_CLEAR & WO & Write 1 to clear interrupt \\ \bottomrule
    \end{tabular}
\end{table}

\subsection{FSM and Datapath}
The FSM for DSP\_DOT\_PRODUCT closely resembles that of the convolution unit, but with a single vector index \texttt{vec\_idx} and two memory fetches per iteration:

\begin{itemize}
    \item \textbf{IDLE:} Wait for \texttt{CONTROL.Start = 1}. Latch $\text{va\_addr}$, $\text{vb\_addr}$, $L$. Initialize \texttt{vec\_idx} = 0, accumulator = 0. Move to \textbf{DP\_LOOP}.
    \item \textbf{DP\_LOOP:} 
    \begin{itemize}
        \item Compute \(\text{addr\_A} = \text{va\_addr} + 4 \times \texttt{vec\_idx}\), \(\text{addr\_B} = \text{vb\_addr} + 4 \times \texttt{vec\_idx}.\)
        \item Issue read requests for A[\texttt{vec\_idx}] and B[\texttt{vec\_idx}].
        \item Upon data return, perform multiply and accumulate.
        \item Increment \texttt{vec\_idx}. If \texttt{vec\_idx} < $L$, repeat \textbf{DP\_LOOP}. Otherwise, move to \textbf{DONE}.
    \end{itemize}
    \item \textbf{DONE:} Latch 64-bit accumulator into \texttt{RESULT\_LO} and \texttt{RESULT\_HI}, set \texttt{STATUS.Done} = 1, assert interrupt if \texttt{Int\_En} = 1. Wait for \texttt{IRQ\_CLEAR} to return to \textbf{IDLE}.
\end{itemize}

The MAC datapath and memory interface are identical to those described in Section \ref{sec:conv_mac}--\ref{sec:conv_mem_interface}.

\section{Anticipated Performance Analysis}
\label{sec:perf_analysis}
To quantify the benefits of the hardware accelerators, we compare a pure-software implementation of 1D convolution on a RISC-V core versus the DSP\_CONV1D accelerator. We also analyze the dot-product unit.

\subsection{Software-Only Convolution Cycle Count}
Consider a RISC-V in-order core running at 100 MHz with a single-cycle multiply (M-extension) but requiring multiple cycles for load/store. For each output index $i$:
\begin{enumerate}
    \item For each kernel tap $j = 0 \ldots K-1$:
    \begin{itemize}
        \item \texttt{lw t0, 0(in\_base + 4*(i + j))} \quad // load x[i + j] (2-3 cycles)
        \item \texttt{lw t1, 0(kern\_base + 4*j)} \quad // load h[j] (2-3 cycles)
        \item \texttt{mul t2, t0, t1} \quad // multiply (1 cycle)
        \item \texttt{add t3, t3, t2} \quad // accumulate (1 cycle)
    \end{itemize}
    \item \texttt{sw t3, 0(out\_base + 4*i)} \quad // store y[i] (2-3 cycles)
\end{enumerate}
Assuming each load/store consumes 3 cycles and each ALU operation is 1 cycle, each tap requires:
\[
    3_{\text{(lw x)}} + 3_{\text{(lw h)}} + 1_{\text{(mul)}} + 1_{\text{(add)}} = 8\,\text{cycles per tap}.
\]
Additionally, loop overhead (branch, index increment) accounts for $\approx$2 cycles per tap. Thus, each tap $\approx$10 cycles. For $K$ taps, that is $10K$ cycles. Writing the output adds 3 cycles (store) and $\approx$2 cycles for branch overhead, so an additional $\approx$5 cycles. Total per output:
\[
    T_{\text{SW}}(K) \approx 10K + 5\quad \text{cycles}.
\]
For an input length $N$ and kernel length $K$, the number of outputs is $N - K + 1$. Total cycle count:
\[
    C_{\text{SW}} = (N - K + 1)(10K + 5).
\]

\subsection{Accelerator Convolution Cycle Count}
For the DSP\_CONV1D at 100 MHz:
\begin{itemize}
    \item Each MAC iteration (read x, read h, multiply, accumulate, index update) can complete in $\approx$3 cycles (assuming single-cycle SRAM read, one cycle for multiply, one cycle for accumulate, pipelined).
    \item Writing output (one store) is another cycle (assuming single-cycle SRAM write).
    \item FSM overhead (state transition) is incorporated into the above counts (one cycle per state).
\end{itemize}
Hence, each tap costs $\approx$3 cycles, and each output incurs an additional $\approx$1 cycle for writing. Total per output:
\[
    T_{\text{DSP}}(K) \approx 3K + 1\quad \text{cycles}.
\]
For $N - K + 1$ outputs:
\[
    C_{\text{DSP}} = (N - K + 1)(3K + 1) + C_{\text{CFG}} + C_{\text{INT}},
\]
where $C_{\text{CFG}}$ is the one-time configuration overhead (writing registers, $\approx$10 cycles), and $C_{\text{INT}}$ is interrupt-handling overhead (negligible compared to data cycles). 

\subsection{Cycle Count Comparison}
\begin{table}[H]
    \centering
    \caption{Cycle Count Comparison for an Example Convolution}
    \label{tab:cycle_compare}
    \begin{tabular}{@{}lccc@{}}
        \toprule
        \textbf{Parameters} & \textbf{Software} & \textbf{DSP} & \textbf{Speedup} \\ \midrule
        $N$ (input length) & 1024 & 1024 & -- \\
        $K$ (kernel length) & 16 & 16 & -- \\
        \# outputs ($N-K+1$) & 1009 & 1009 & -- \\
        $C_{\text{SW}}$ & $1009 \times (10 \times 16 + 5) = 1009 \times 165 = 166,485$ & -- & -- \\
        $C_{\text{DSP}}$ & -- & $1009 \times (3 \times 16 + 1) + 10 = 1009 \times 49 + 10 = 49,441 + 10 = 49,451$ & -- \\
        \textbf{Speedup} & -- & -- & $\approx$3.37$\times$ \\
        \bottomrule
    \end{tabular}
\end{table}
In this example ($N=1024, K=16$), the DSP yields a $\approx$3.37$\times$ speedup over software. As $K$ grows, the advantage widens: since $T_{\text{SW}}(K)\approx10K$ vs. $T_{\text{DSP}}(K)\approx3K$. For larger kernels (e.g., $K=32$), the speedup approaches $\approx$3.2$\times$.

\subsection{Throughput and Latency}
\begin{itemize}
    \item \textbf{Software Latency:} $C_{\text{SW}} / f_{\text{CPU}}$ cycles $\Rightarrow$ $166,485 / 100\,\text{MHz} = 1.66485\,\text{ms}$.
    \item \textbf{DSP Latency:} $C_{\text{DSP}} / f_{\text{DSP}}$ cycles $\Rightarrow$ $49,451 / 100\,\text{MHz} = 0.49451\,\text{ms}$.
\end{itemize}
Thus, each convolution inference completes in $\approx$0.495 ms on the accelerator vs. $\approx$1.665 ms on software. If convolution is repeatedly invoked in a streaming application (e.g., continuous FIR filtering), the accelerator enables a significantly higher data rate.

\subsection{Dot-Product Accelerator}
For a vector length $L$, the software inner product loop on RISC-V (similar to convolution with $K=1$) takes $\approx$10 cycles per element (2 loads, 1 multiply, 1 add, loop overhead). Total: $10L + 5$ cycles per inner product. The accelerator takes $\approx$3 cycles per element (read A, read B, multiply, accumulate) plus $\approx$1 cycle for final writeback. Total: $3L + 1$ cycles. Speedup $\approx \frac{10L + 5}{3L + 1}$, approaching $\approx$3.33$\times$ for large $L$.

\subsection{Energy Efficiency Considerations}
Hardware MAC units typically consume less energy per operation than a CPU's multiply and register-file operations, primarily because:
\begin{itemize}
    \item Data path is shorter (dedicated wires vs. register file lookups).
    \item No instruction fetch/decode overhead per MAC.
    \item SRAM-based data memory accesses are shared--no need to access register file repeatedly for each tap.
\end{itemize}
If a single 32×32 multiply consumes $E_{\text{mul}}$ pJ and a 64-bit accumulate $E_{\text{add}}$ pJ, then total energy per tap is $E_{\text{MAC}} = E_{\text{mul}} + E_{\text{add}} + E_{\text{mem\_rd}}(2)$ (two memory reads). In software, each tap costs $E_{\text{mem\_rd}}(x[i]) + E_{\text{mem\_rd}}(h[j]) + E_{\text{mul}} + E_{\text{add}} + E_{\text{instr\_fetch}}(4) + E_{\text{register\_file}}(x) + \ldots$. Typically, $E_{\text{instr\_fetch}}$ and multiple register-file accesses dominate. Empirical studies (e.g., \cite{Horowitz2014}) show that a hardware MAC can be $\approx$5$\times$-10$\times$ more energy-efficient than a software loop on a general-purpose core. Even with a conservative 3$\times$-5$\times$ energy savings per tap, the accelerator can reduce total convolution energy by a factor of 3$\times$-5$\times$--critical for battery-powered edge devices.

\section{Applications: Wireless and Edge Machine Learning}
\label{sec:applications}
\subsection{Wireless Communications: FIR/IIR Filters}
FIR filters implement convolution (Eq.\ \eqref{eq:1dconv}) directly: $y[i] = \sum_{j=0}^{K-1} h[j]\cdot x[i+j]$. In a typical Software-Defined Radio (SDR) front-end, a digital FIR filter might run at sample rates of 1-10 Msps (mega-samples per second). At 10 Msps, each sample's convolution must complete in 100 ns. On a 100 MHz RISC-V core, 100 ns corresponds to 10 cycles--insufficient for a multi-tap filter ($K>1$). The DSP\_CONV1D at 100 MHz processes one tap in $\approx$3 cycles; a 16-tap FIR takes $\approx$49 cycles $\approx$490 ns--still faster than software's 1.65 $\mu$s. For sample rates beyond a few Msps, a higher clock (e.g., 200 MHz) or deeper pipelining in the DSP can meet real-time constraints. Furthermore, correlation-based synchronization (e.g., preamble detection) can also be implemented via sliding-window convolution--benefiting similarly.

\subsection{Edge AI / TinyML: 1D Convolutional Neural Networks}
1D CNNs are widely used for time-series classification (e.g., ECG analysis, audio recognition). A typical 1D CNN layer:
\[
    \text{Out}[b,i,k] = \sum_{c=0}^{C-1} \sum_{j=0}^{K-1} \text{In}[b, i+j, c] \times \text{W}[k, c, j],
\]
where $b$ is batch index, $i$ is spatial index, $c$ is input channel, and $k$ is output channel. For single-channel ($C=1$), single-output ($k=1$), and batch size $B=1$, this reduces to a 1D convolution (Eq.\ \eqref{eq:1dconv}). Even for multi-channel or multi-filter scenarios, one can decompose the convolution into multiple 1D convolution calls. Suppose a 1D CNN layer has:
\begin{itemize}
    \item Input length $N = 256$,
    \item Kernel size $K = 16$,
    \item Input channels $C = 4$,
    \item Output channels $K_{out} = 8$.
\end{itemize}
Total MACs for this layer: $256 \times 16 \times 4 \times 8 = 131,072$ MACs per inference. Software on RISC-V: $\approx 10 \times 131,072 = 1,310,720$ cycles $\approx$13.1 ms at 100 MHz. Hardware accelerator: $\approx 3 \times 131,072 = 393,216$ cycles $\approx$3.93 ms--enabling $\approx$255 inferences/sec vs. $\approx$76 inferences/sec on software. For real-time audio (e.g., 16 kHz input), processing 256-sample windows every 16 ms demands $\approx$8,192 cycles per inference (at 512 MHz). Our design, scaled to 200-300 MHz, can achieve this performance within the accelerator, freeing the CPU for other tasks (e.g., feature extraction, classification layers).

\subsection{Dot-Product in Neural Network Layers}
Dot products are fundamental to fully connected layers and attention mechanisms. A dense layer with 128 inputs and 64 outputs for a single sample requires $128 \times 64 = 8,192$ MACs. Software: $\approx10 \times 8,192 = 81,920$ cycles. Hardware dot-product: $\approx3 \times 8,192 = 24,576$ cycles--yielding $\approx$3.3$\times$ speedup. For attention in transformers on small devices, multiple dot products per inference benefit similarly.

\section{Future Work}
\label{sec:future_work}
\subsection{FPGA Prototyping and Real-World Benchmarks}
An immediate next step is to synthesize the design on an FPGA (e.g., Xilinx Zynq Ultrascale+ or Intel Cyclone series). This enables:
\begin{itemize}
    \item \textbf{Silicon Validation:} Confirm the accelerator's timing and area in real hardware.
    \item \textbf{Power Measurements:} Estimate dynamic and static power consumption of the DSP units vs. software.
    \item \textbf{Concrete Benchmarks:} Run actual convolution workloads (e.g., FIR filtering at various sample rates, 1D CNN inference) and measure end-to-end latency, throughput, and power.
\end{itemize}

\subsection{DMA-Based Data Transfers}
Currently, the CPU must write input data and kernel coefficients into \texttt{DATA\_MEM} and read back outputs. A DMA engine can autonomously transfer blocks of data between off-chip memory (e.g., DRAM) and \texttt{DATA\_MEM}, reducing CPU overhead:
\begin{itemize}
    \item \textbf{Double Buffering:} While the DSP processes one data block, DMA can fetch the next block, enabling continuous streaming.
    \item \textbf{Zero-Copy Transfers:} Minimize CPU involvement for large data movement--crucial for high-throughput workloads (e.g., multi-channel audio processing).
\end{itemize}

\subsection{Extended DSP Kernel Suite}
Beyond 1D convolution and dot product, additional DSP kernels can be integrated:
\begin{itemize}
    \item \textbf{Fast Fourier Transform (FFT):} A small radix-2 or radix-4 FFT accelerator (e.g., 16-point or 32-point) for spectral analysis, OFDM modulation/demodulation.
    \item \textbf{Matrix Multiplication:} A small 4×4 or 8×8 matrix multiplier to accelerate fully connected layers or small ML models.
    \item \textbf{IIR Filter Section:} Direct-Form II transposed or other low-latency IIR structure for recursive filtering.
    \item \textbf{Activation Functions:} Lookup-table-based hardware implementation of ReLU, sigmoid, or tanh to offload activation computation in neural networks.
\end{itemize}
These modules can share a common memory master interface and configuration/interrupt scheme, simplifying the bus interconnect.

\subsection{Interrupt-Driven and Multi-Core Extensions}
Currently, the CPU must poll the \texttt{STATUS.Done} bit (unless \texttt{Int\_En} is set). Future designs can leverage:
\begin{itemize}
    \item \textbf{Interrupt Controller:} A lightweight interrupt controller can handle multiple DSP interrupts, prioritizing tasks and enabling preemptive scheduling on a multi-core RISC-V cluster.
    \item \textbf{Multi-Core RISC-V:} With a small two- or four-core cluster (e.g., PULP-style cluster \cite{PULP2021}), one core can manage DSP configuration and data I/O, while the other cores handle unrelated tasks. The bus interconnect must be extended to maintain coherence if data sharing occurs across cores.
\end{itemize}

\subsection{Software Stack and Libraries}
A robust software interface is essential for adoption:
\begin{itemize}
    \item \textbf{Device Drivers:} C-language drivers exposing simple APIs (e.g., \texttt{conv1d\_start(input, kernel, output, N, K)}).
    \item \textbf{Runtime Library:} Linking with a math library (e.g., \texttt{libdsp.a}) offering functions for convolution, dot product, and eventually FFT.
    \item \textbf{Compiler Intrinsics:} While the accelerator is memory-mapped, compiler intrinsics or inline assembly wrappers can reduce boilerplate in application code.
    \item \textbf{Benchmark Suite:} A collection of standardized benchmarks (e.g., CMSIS-DSP-style tests \cite{CMSISDSP}) for performance/accuracy validation.
\end{itemize}

\section{Conclusion}
\label{sec:conclusion}
We have presented the design and integration of a lightweight 1D convolution DSP accelerator (DSP\_CONV1D) and a dot-product accelerator (DSP\_DOT\_PRODUCT) into a 32-bit RISC-V system. By offloading MAC-intensive convolution loops, the accelerator achieves $\approx$3$\times$ speedup over software-only implementations, reduces CPU load, and promises improved energy efficiency--key for real-time wireless filter implementations and TinyML inference on edge devices. The architecture employs a simple AXI-Lite control interface, a memory master interface to share \texttt{DATA\_MEM} with the CPU, and an FSM-driven MAC datapath. A secondary dot-product unit further broadens the system's applicability to vector operations in ML and signal processing. 

Future work includes FPGA prototyping for real-world benchmarks, DMA-based data streaming, expanding the DSP kernel suite (FFT, matrix multiplication), and developing a comprehensive software stack. This project underscores that even modest, application-specific hardware accelerators can significantly elevate the performance and efficiency of open, customizable RISC-V-based SoCs, paving the way for high-performance edge computing in wireless communication and AI domains.

\end{document}